\documentclass[wrr]{agu2001}

\usepackage[dvips]{graphicx}

\authorrunninghead{NOVEMBER}
\titlerunninghead{HILLSLOPE MOISTURE CONTENT}

\authoraddr{L. J. November, La Luz Physics, La Luz NM 88337-0217 USA
({\tt laluzphys@yahoo.com})}

\begin{document}

\title{Dependence of hillslope moisture content on downhill
saturation}

\author{L. J. November}
\affil{La Luz Physics, La Luz NM 88337-0217 USA\\
January 16, 2008}

\begin{abstract}
 We derive steady equilibria for lateral downslope moisture flow in
an idealized thin closed layer as a solution to the 1D Richards'
Equation.  The equilibria are determined by two free parameters: the
downslope flux and a boundary condition. Solutions exhibit a constant
downslope flow speed and moisture content for the constant
equilibrium flux, which is the product of the two.  However where an
isolated zone of fixed saturation degree exists representing a
boundary condition, the flow speed immediately upslope is reduced and
the moisture content correspondingly increased to preserve the
constant equilibrium flux. The capillary head jump at the saturated
zone produces a blockage that gives a high moisture content back
upslope through a pooling distance determined by the equilibrium
condition that the downslope flux is constant.  In our numerical
integrations, the vertically projected pooling height is more than 10
km for a fully saturated zone in mixed silty or clay soils, but
decreases by about an order of magnitude with every 10\% decrease in
the boundary-zone saturation degree.  The drying of downhill
saturated zones with the increased speed of mountain moisture outflow
and corresponding decreased mountain moisture content gives a viable
explanation for the mysterious $~69\%$ unaccounted drop seen in the
spring outflow in the La Luz / Fresnal Watershed at Alamogordo's
upstream spring-box diversions in the semiarid southeastern New
Mexico USA. 

\noindent
[agu index terms: hydrology (1800), groundwater hydrology (1829),
groundwater transport (1832), streamflow (1860), soil moisture (1866),
vadose zone (1875)]
 \end{abstract}

\begin{article}
\section{Introduction}
\label{s:intro}

\begin{figure}
\centering
\includegraphics[width=8.5cm]{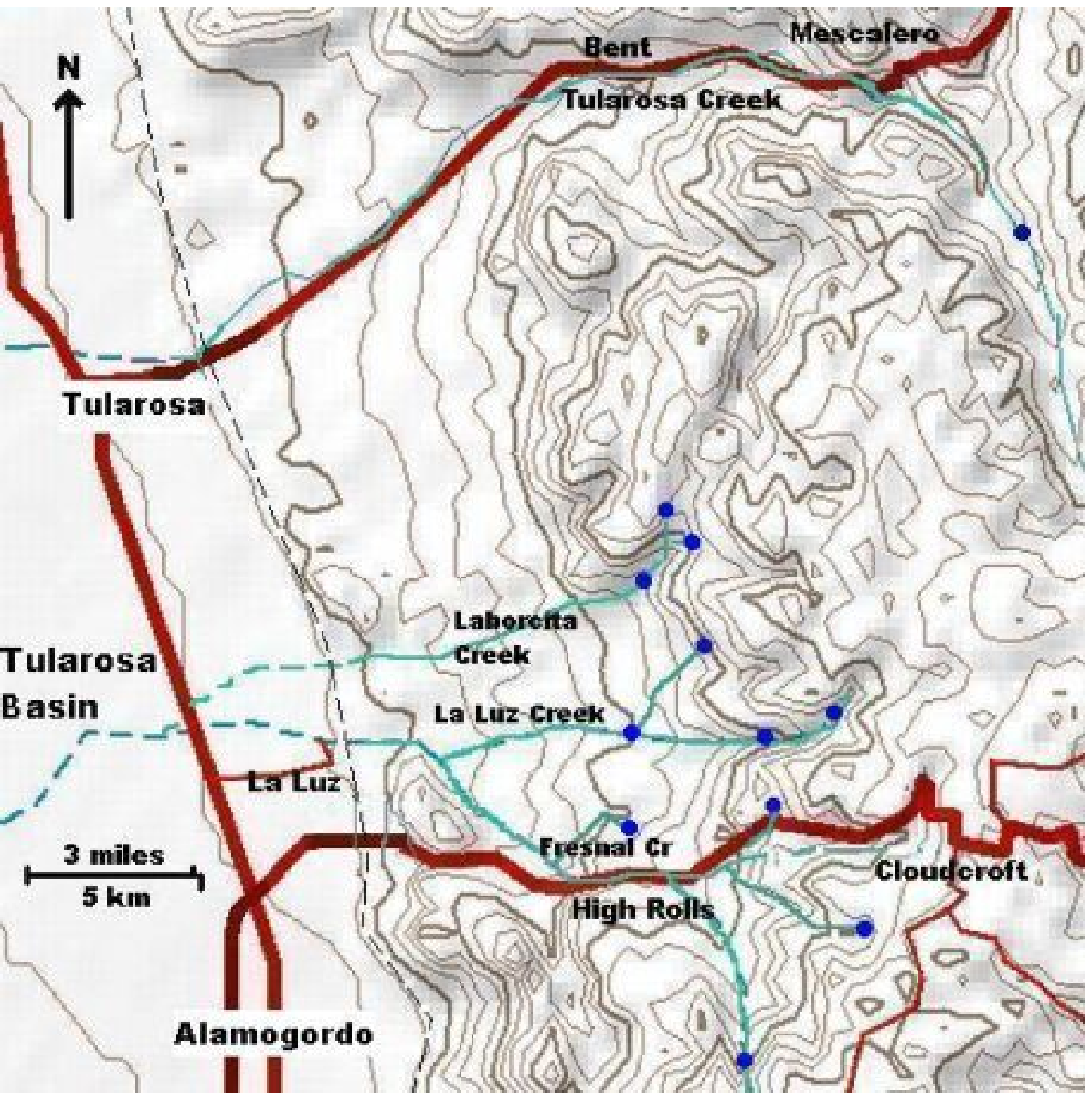}
 \caption{Topo Map of western Sacramento Mountains NM USA with
contour interval 250 ft (76.2 m) and heavy-contour interval 1000 ft
(305 m) from 5000 ft (1524 m) to 9000 ft (2745 m) above sea level;
the total vertical range is 4200 ft (1280 m); blue dots are the main
springs; normal streams are shown as solid blue and normally dry
streambeds as dashed blue; the black dashed line denotes the foot of
the mountains; major roads are shown as red.}
 \label{f:lltopo}
\end{figure}

Largely with the hope of increasing the captured outflow of the La
Luz / Fresnal Canyon stream system, the Town of Alamogordo New Mexico
substantially repaired and moved its collection system between
1982-1997, to divert nearly all of the streamflow at or just below
the mountain spring sources.  The locale is shown on the topo map in
Figure \ref{f:lltopo} with the main springs indicated.  By
transporting the water in pipes through the lower canyons of the
western Sacramento Mountains they minimize evapotranspiration losses
and avoid possible losses due to unappropriated human usage, which is
believed to have been a problem historically.

\begin{figure}
\centering
\includegraphics[width=8.5cm,trim=10 50 -10 -50]{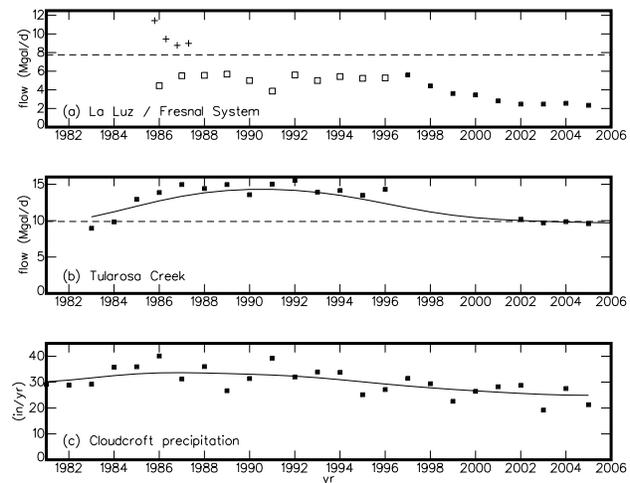}
 \caption{(a) La Luz / Fresnal System partial- ({\it empty boxes}),
and full- diversions ({\it filled boxes}), historic average baseflow
({\it dashed line}), and medians of daily flows ({\it plus signs});
(b) Tularosa Creek yearly medians ({\it boxes}), smoothed with a
4-year FWHM Gaussian ({\it line}), and historic average baseflow
({\it dashed}); (c) Cloudcroft precipitation yearly averages and
4-year smoothed ({\it line}).}
 \label{f:laluz}
\end{figure}

The 100-year-span USGS-monitored baseflow at the foot of the
mountains near the village of La Luz was 7.8 million gallons per day
(Mgal/d) (342 l/s), shown as a dashed line in Figure \ref{f:laluz}a
(USGS reports summarized in Table E-6 of \citealt{RCD2002}), but in
the mid 1980s the streamflow ranged up to about 11.5 Mgal/d (504 l/s)
in the median of the daily averages selected from two-month periods
when the flows were consistently high (plus signs, from New Mexico
Water Resources Data, USGS yearly reports).  The La Luz / Fresnal
streamflow was quite variable since the USGS monitoring resumed after
1980 due to the intermittent operation of Alamogordo's diversions,
but during 1985-87 the La Luz spring capture boxes were not operated
and the streams allowed to flow fully during the transition to the
new system. A consistently high flow during the transition period
indicates that Alamogordo was not diverting the water elsewhere in
the system and allows us to obtain a best minimum estimate for the
full natural streamflow.  The empty boxes in Figure \ref{f:laluz}a
show the average metered pipe flow made during periods when some
streamflow was allowed and give a lower bound on the actual system
outflow; the filled boxes are for periods when the entire flow was
piped and streams dryed, and represent essentially the full system
outflow (New Mexico Office of the State Engineer, unpublished data,
2006; Town of Alamogordo, unpublished data, 2007).  The data point
rms accuracy is limited by the inherent device precision, which for
Alamogordo's pipe-flow measurements is 2\%, and for the USGS
streamflow measurements is estimated to be 5\%, the minimum noise for
the daily averages obtained with their system.

However after the upstream diversion relocation was complete, rather
than increase, the total outflow of the system dropped to less than
2.5 Mgal/d (110 l/s) by 2002. The drop represents a huge decrease from
about 11.5 Mgal/d (504 l/s) to less than 2.5 Mgal/d (110 l/s) or a
loss of about 78\%!  During the same 1986-2002 period, the average
precipitation in the Cloudcroft and mountain source area decreased by
about 25\%, as shown by the smoothed curve in Figure \ref{f:laluz}c
(\citealt{wrcc2006}).  Similar Tularosa Creek showed a larger 30\%
drop during the period suggesting a streamflow overresponse to the
precipitation change and implying a reduced projected La Luz /
Fresnal flow for the dry period of about 8 Mgal/d (351 l/s).  Thus
without other significant appropriations or diversions, we estimate a
drop in the total system outflow from about 8 Mgal/d (351 l/s) to
less than 2.5 Mgal/d (110 l/s) meaning that about 69\% of the normal
system outflow cannot be accounted for since the mid-1980's.  The
unaccounted drop is probably conservative since downstream losses are
not considered in comparing upstream spring outflow to downstream
flow.  However small secondary sources are also known to exist, which
act oppositely, adding to the downstream flow.

As a baseline, the USGS daily record from Bent NM shows that Tularosa
Creek about 10 miles (16 km) to the north, which has remained intact,
has changed consistently with the mountain precipitation with a
3.76 year delayed response, evident in comparing the smoothed data
lines between Figures \ref{f:laluz}b and \ref{f:laluz}c (New Mexico
Water Resources Data, USGS yearly reports).  Since 1998, Cloudcroft
and its surroundings above the La Luz / Fresnal Watershed have
experienced considerable drying of wells and springs, with a forest
mortality consistent with a reduced mountain moisture content,
whereas no similar depletions or its effects seem evident in
Mescalero and the mountains above Tularosa Creek.  Laborcita Creek,
which has a much smaller normal outflow, was not monitored
consistently during the period. 

The drop in outflow seen in the La Luz / Fresnal Stream System is
regionally localized, appears to be historically unprecedented, and
has been persisting after 2002, so most common effects can be ruled
out.  For example, precipitation patterns, snowpack variation or
storm distribution and frequency, have larger spatial scale and are
historically recurring.  A similarly large drop in outflow was not
seen in the other major stream systems in the locale, and an
unusually large drop did not occur in the La Luz / Fresnal Stream
System following the 1970s dry period.  The loss is greater than
all the other appropriations in the system combined and much larger
than the 0.56 Mgal/d (24.5 l/s) increase in appropriations for the
1980-2000 period, which was for domestic wells mainly located below
the Alamogordo diversions (from New Mexico Office of the State
Engineer, WATERS database www.ose.state.nm.us, 2007).  Other water
users in the system have experienced similar drops in their spring
flow and drying of their wells during the period too. Thus a
significant impact due to other human usage seems precluded.

The steady and precipitous drop with approximately the same e-folding
time as the delayed response in the Tularosa Creek system, and
apparently nearly uniform loss in all of the springs after 1997 when
Alamogordo started taking all of the streamflow, suggests a
hydrological phenomenon.  Rick Warnock, president of the Sacramento
Mountain Watershed Restoration Corporation, has maintained that the
lower-canyon drying must be the main cause of the mountain water
depletion (\citealt{WarnockR2006a}).  In one study in another system,
diurnal and bi-annual moisture variations are found to be strongly
linked to changing downstream river flow
(\citealt{Waichler+Yabusaki2005}).

Such a large anomalous event gives unique opportunity for a critical
hydrological test:  Is it possible that downstream drying could cause
upslope moisture depletion?  A downstream influence on upslope
conditions may be possible if unsaturated lateral downslope moisture
dynamics plays a dominant role in controlling the water distribution
throughout the system.  The relative importance and interrelation of
saturated/unsaturated dynamics in downslope flow has been a matter of
study and discussion (\citealt{Freeze+Witherspoon1966};
\citealt{JacksonCR1992}).  Numerous measurements indicate direct 3D
unsaturated near-surface moisture flow, suggesting that a significant
downslope flow component may remain as unsaturated (e.g.
\citealt{Weyman1973}; \citealt{Nieber+Walter1981};
\citealt{Miyazaki1988}; \citealt{Sinai+Dirksen2006}).  A largely
unsaturated downslope flow may be suggested by the fact that the
streamflow in these systems is never larger than about 15\% of the
total precipitation inflow.

We seek guidance on the gross characteristics of large-scale mountain
moisture storage dynamics outlined here.  The geological and
topographical constraints suggest that moisture leaves the mountains
traveling through the lower canyons mainly as a lateral downslope
flow in a thin layer being prevented from seeping down very far
because of underlying impervious layers.  The Richards' Equation
describes general moisture flow in mixed saturated/unsaturated
ground-water systems giving a correct upper vadose zone, capillary
fringe, water table, and lower saturated zone in simulations, though
its multidimensional form is difficult to model due to its nonlinear
properties (\citealt{Fipps+Skaggs1989}; \citealt{Fiori+Russo2007}).

One published solution to the Richards' Equation for general
downslope flow in a thin layer assumes a constant downslope flow
speed and moisture content (\citealt{Philip1991a}), while others show
pooling at the bottom of the hillslope (\citealt{Sloan+Moore1984};
\citealt{Hurley+Pantelis1985}; \citealt{Stagnitti+++1986};
\citealt{Steenhuis+++1999}).  As we show in this study, flow
solutions of both these types are obtained depending upon the choice
of downhill boundary condition.  The boundary condition may be
affected by streamflow maintaining a downhill saturated zone.  Where
a high downhill saturation degree is maintained, moisture backs
upslope through a pooling distance that depends upon the soil type. 
Alamogordo's diversions near the canyon spring sources dry the
lower-canyon stream system, which must dry the lower saturated zones
and change the downhill boundary condition. Such drying leads to a
higher-speed lateral downslope moisture flow into the open alluvial
basin below with a decreased moisture content back upslope as we
describe.  The reduced moisture content can affect the nearby spring
outflow as well as possibly the relative mountain moisture content
above.

\section{1D Steady Moisture Flow}
\label{s:1Dsteady}

\begin{figure}
\centering
\includegraphics[width=8.5cm,trim=5 20 -5 -20]{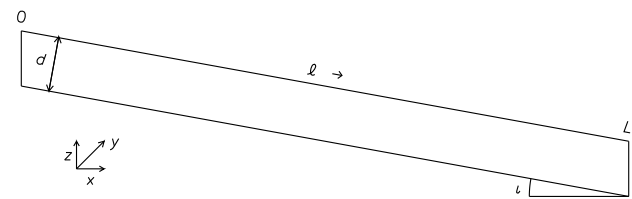}
 \caption{Idealized 1D friction flow down a uniformly filled closed
flow layer of constant thickness $d$ and length $L$, inclined at
$\iota$ from the horizontal $x$, repeated infinitely into the plane
of the drawing in $y$; $z$ is the vertical and $\ell$ the downslope
coordinate, which ranges from 0 to $L$.}
 \label{f:drawing}
\end{figure}

Friction flow down a uniformly filled closed layer of constant
inclination $\iota$ and thickness $d$ is an idealization of the flow
of mountain moisture to a lower basin outlet over a distance $L$, as
illustrated in Figure \ref{f:drawing}.  By Darcy's rule, the flux (or
specific flux) ${\vec q}$ with units of speed (L/T) is always away
from the pressure head (Chapter 5, \citealt{Bear1988})
 \begin{equation}
{\vec q}=-K(S)\left({\vec\nabla}\psi(S)+{\vec\nabla}z\right),
\label{e:flux}
\end{equation}
 written for isotropic conditions, where the scalar hydraulic
conductivity $K(S)$ with units of speed (L/T) and the capillary
pressure head $\psi(S)$ with units of length (L) are functions of the
saturation (or saturation degree) $S$ and the soil properties;
${\vec\nabla}z$ represents the gradient of the gravitational pressure
head.  The saturation degree $S$ ranges from 0 to 1 in proportion to
the fractional water filling of the available spaces between soil
grains, with $S=1$ indicating fully saturated. The flux ${\vec q}$ is
proportional to the flow speed ${\vec V}$
 \begin{equation}
{\vec q}=\theta{\vec V}=n S{\vec V},
\label{e:Vdef}
\end{equation}
 where $\theta=nS$ defines the moisture content (Section 9.4.4
\citealt{Bear1988}), and $n$ is the porosity, the fractional volume
of open space between soil grains.  Though in our mathematical
formulation a spatially varying porosity is allowed, in general
discussions concerning the downslope flow speed and moisture content,
a uniform porosity is assumed for simplicity.

The conservation of moisture is represented by the incompressible
form of the continuity equation (Section 9.4, \citealt{Bear1988})
 \begin{equation}
n{\partial S\over\partial t}=-{\vec\nabla}\cdot{\vec q},
\label{e:cont}
\end{equation}
 assuming that the medium is nondeformable or the porosity $n$ not
temporally varying.  We have neglected other terms commonly
introduced into the continuity equation representing precipitation
influx, evapotranspiration loss, or seepage out the bottom below the
hilltop entry point at $\ell=0$.  While such effects might be
included, they should not alter the essential physical properties of
the solutions, as we discuss further in Section \ref{s:3D}.

Combining (\ref{e:flux}) and (\ref{e:cont}) yields the Richards'
general flow equation (\citealt{Richards1931})
 \begin{equation}
n{\partial S\over\partial t}={\vec\nabla}\cdot\left(K(S)
\left({d\psi\over dS}{\vec\nabla}S+{\vec\nabla}z\right)\right),
\label{e:flow}
\end{equation}
 having expanded the gradient of the capillary head with its implicit
spatial dependence ${\vec\nabla}\psi(S({\vec x}))=
(d\psi/dS){\vec\nabla}S({\vec x})$.

We are primarily interested in the lateral downslope component to the
flux $q_\ell$, which satisfies the $\ell$ vector-element equation of
(\ref{e:flux})
 \begin{equation}
q_\ell=K(S)\left(\sin\iota-{d\psi\over dS}{\partial S\over\partial\ell}\right),
\label{e:1Dflux}
\end{equation}
 using $\partial z/\partial\ell=-\sin\iota$, for the hillslope
inclination $\iota$.

We adopt the 1D approximation as justified and used in other similar
studies (\citealt{Sloan+Moore1984}; \citealt{Hurley+Pantelis1985};
\citealt{Stagnitti+++1986}; \citealt{Steenhuis+++1999}), supposing
that moisture changes normal to the surface are small compared to
lateral changes down the flow layer, which should be the case when
the idealized flow layer of Figure \ref{f:drawing} is thin.  The 1D
flow vector can be written ${\vec q}= q_\ell {\vec\nabla}\ell$, and
with variations only in the downslope direction $\ell$, the flux and
saturation written $q_\ell=q_\ell(\ell,t)$ and $S=S(\ell,t)$. The
continuity equation (\ref{e:cont}) in the 1D approximation is written
 \begin{equation}
n{\partial S\over\partial t}=-{\partial q_\ell\over d\ell}.
\label{e:1Dcont}
\end{equation}

Steady equilibrium solutions are obtained with $\partial S/\partial
t=0$ in (\ref{e:1Dcont}), which gives a constant flux $q_\ell$
downslope in $\ell$.  A constant saturation down the flow layer
$S(\ell)=\hat{S}$ for the reference saturation degree $\hat{S}$ gives
the constant flux $q_\ell$ and represents one steady solution to
(\ref{e:1Dflux}), written
 \begin{equation}
q_\ell= K(\hat{S})\sin\iota.
\label{e:hatS}
\end{equation}
 However general steady equilibria with a varying $S(\ell)$ and
nonzero derivative $\partial S/\partial\ell$ in (\ref{e:1Dflux}) also
exist. Solving for the derivative gives the governing differential
equation for saturation degree as a function of position down the
flow layer $S(\ell)$
 \begin{equation}
{\partial S\over\partial\ell}=
\sin\iota\ \left({d\psi\over dS}\right)^{-1}\left(1-{K(\hat{S})\over K(S)}\right),
\label{e:1DSa}
\end{equation}
 now using the reference saturation degree $\hat{S}$ as a problem
constant to represent the constant flux $q_\ell$ defined by
(\ref{e:hatS}).  Since $\partial z=-\partial\ell\sin\iota$,
dependencies on the inclination angle simply project, and the
equation exhibits universal solutions $S(z)$ as functions of the
vertical coordinate $z$ measured positively upward and independent of
the inclination angle $\iota$
 \begin{equation}
{\partial S\over\partial z}=
\left({d\psi\over dS}\right)^{-1}\left({K(\hat{S})\over K(S)}-1\right).
\label{e:1DSb}
\end{equation}
 Solutions $S(z)$ to the first-order differential equation are
describable by one additional free parameter besides the constant
flux $q_\ell$ or reference saturation $\hat{S}$, a boundary value
$S(\ell)$ at some downslope position $\ell$, which may represent the
saturation in a fixed constant boundary zone.  The boundary condition
only makes physical sense as a downhill condition given the form of
the solutions found in Section \ref{s:num}, as we discuss further in
Section \ref{s:phys}.

We adopt the van Genuchten material functions for $K(S)$ and
$\psi(S)$ (\citealt{vanGenuchten1980};
\citealt{vanGenuchten+Nielsen1985}), written in terms of the
effective saturation degree $S_{\rm e}(S)=(S-S_{\rm r})/(1-S_{\rm
r})$, for $S_{\rm r}$ a small retained moisture saturation degree
 \begin{equation}
K(S)=K_{\rm sat} S_{\rm e}(S)^{1\over2}\left(1-\left(1-S_{\rm e}(S)^{1\over\eta}\right)^\eta\right)^2,
\label{e:K}
\end{equation}
 \begin{equation}
\psi(S)=\psi_0\left({1\over S_{\rm e}(S)^{1\over\eta}}-1\right)^{1-\eta},
\label{e:psi}
\end{equation}
 where $K_{\rm sat}$, the saturated hydraulic conductivity, $\eta$,
and $\psi_0$ are material constants.  The retained moisture
saturation degree $S_{\rm r}$ is a small zero offset for the
saturation curves that depends upon the soil type.  For our
calculations, we suppose uniform soil properties with constant
material parameters $K_{\rm sat}$, $\eta$, $\psi_0$, and $S_{\rm r}$,
and use $S_{\rm r}=0.1$ .

\begin{figure}
\centering
\includegraphics[width=8.5cm,trim=10 50 -10 -50]{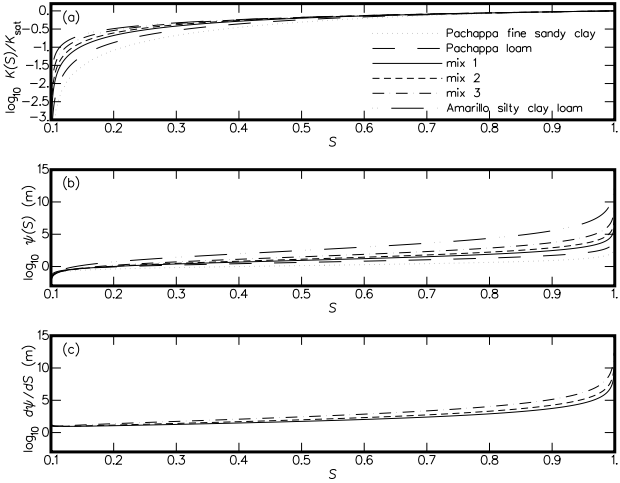}
 \caption{Van Genuchten material functions with saturation degree
$S$: (a) relative hydraulic conductivity $K(S)/K_{\rm sat}$; (b)
capillary head $\psi(S)$ (m); and (c) derivative $d\psi/dS$ (m), for:
Pachappa fine sandy clay ($\eta=$ 1.860, $\psi_0=$ 0.80 m, {\it
dotted}), Pachappa loam ($\eta=$ 2.195, $\psi_0=$ 1.65 m, {\it long
dash}), hypothetical mix 1 ($\eta=$ 2.7, $\psi_0=$ 1.5 m, {\it
solid}), mix 2 ($\eta=$ 3., $\psi_0=$ 1.3 m, {\it dashed}), mix 3
($\eta=$ 3.5, $\psi_0=$ 1.1 m {\it dash dot}), and Amarillo silty
clay loam ($\eta=$ 4.510, $\psi_0=$ 1.015 m, {\it long-dash dot dot})
all with $S_{\rm r}=0.1$.}
 \label{f:vG}
\end{figure}

The Van Genuchten formulation is empirical, but is a good
approximation for soil measurements as described by Assouline and
collaborators in their fully developed theoretical formulation of the
problem.  \citet{Assouline+Tartakovsky2001} give best fits for the
van Genuchten coefficients for different materials in their Table 2.
Figure \ref{f:vG} illustrates the van Genuchten material functions
for various soil samples.  The singular behavior in the derivative
$d\psi/dS$ approaching full saturation $S\rightarrow 1$ leads to a
very small saturation change with height near full saturation in
(\ref{e:1DSb}) and solutions with large pooling heights. The La Luz /
Fresnal Watershed consists of mixed soils and should correspond to an
intermediate silty soil between coarse sand $\eta<2.2$ and dense clay
$\eta>3.5$, taken between Pachappa Loam and hypothetical mix 2.

Solutions to (\ref{e:1DSb}) are independent of the saturated
hydraulic conductivity $K_{\rm sat}$, which divides out in the ratio
$K(\hat{S})/K(S)$.  Such independence gives the equilibria certain
robustness as $K_{\rm sat}$ is the least certain material constant
due to the inhomogeneous character of real ground material, which can
allow some relatively high-speed horizontal flow in cracks and open
zones (\citealt{Sigda+Wilson2003}).  It ranged from about 0.01 to 3
m/d in field tests (from the averages in Table 1 of
\citealt{XiangJ+++1997}), but to much larger values $>50$ m/d with a
median of 7 m/d for a sample of wells from draw-down measurements
made in the Tularosa Basin (\citealt{Orr+Myers1986}).  We suppose
$K_{\rm sat}=7$ m/d for estimating downslope flow speeds.  The choice
gives a saturated flow speed $V_{\rm sat}= K_{\rm sat}\sin\iota/n=$
5.0 m/d, combining (\ref{e:hatS}) and (\ref{e:Vdef}) for $S=1$, with
$\iota= 7^\circ$ as the half-mountain-height inclination, and with
the choice $n=0.17$ for the porosity of poorly sorted soils (Section
2.5.2, \citealt{Bear1988}).  For an approximate 7 km mid-mountain
characteristic downslope spatial scale for upper La Luz Creek from
Figure \ref{f:lltopo}, we obtain a saturated-flow timescale of 3.8
years for the system, consistent with the e-folding time for the drop
after 1997 seen in Figure \ref{f:laluz} or the delayed response for
Tularosa Creek with changing precipitation.

\section{Numerical Solutions}
\label{s:num}

\begin{figure}
\centering
\includegraphics[width=8.5cm,trim=10 50 -10 -50]{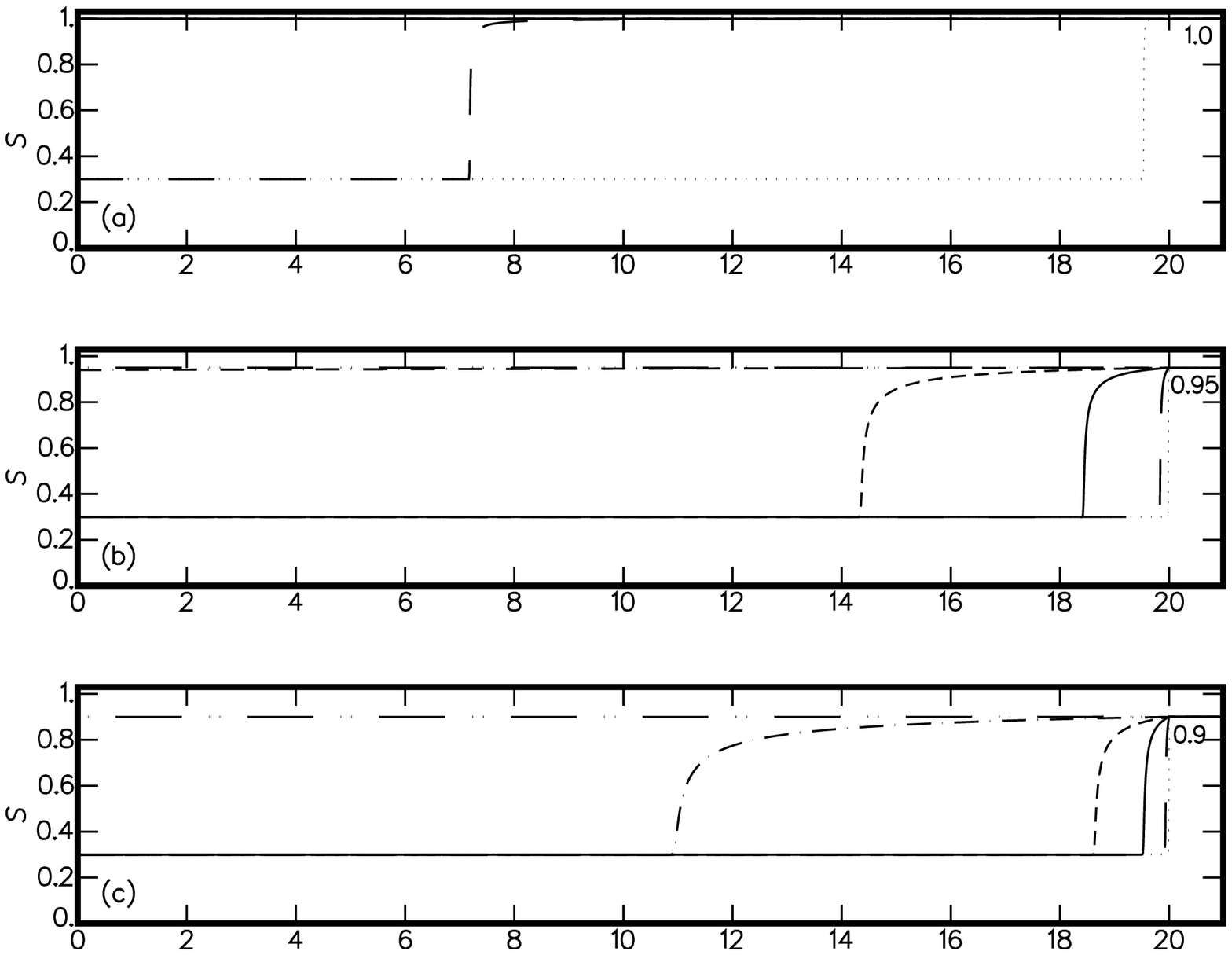}
\includegraphics[width=8.5cm,trim=10 50 -10 -50]{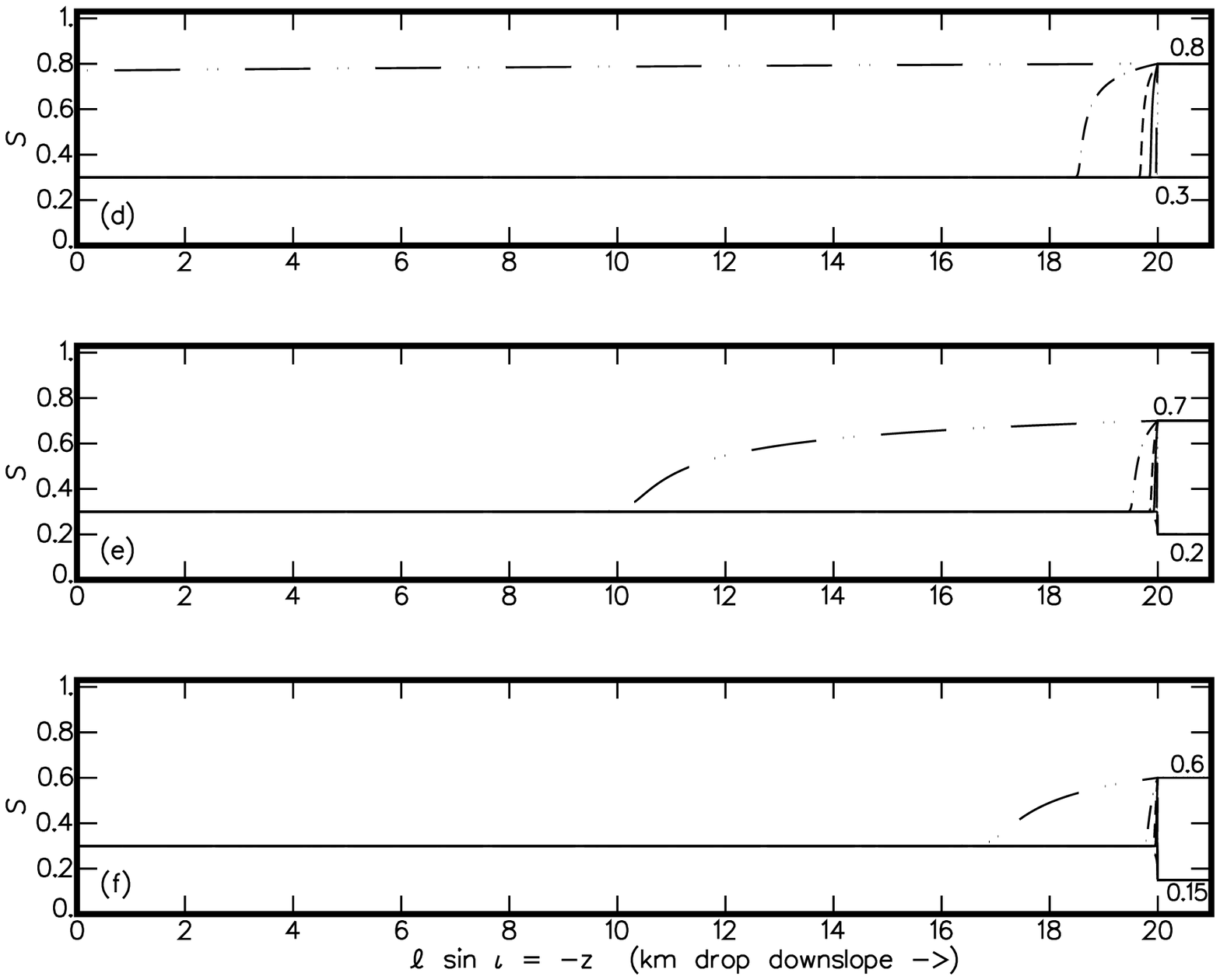}
 \caption{Saturation degree $S$ as a function of vertical drop
$\ell\sin\iota=-z$ (km) down the flow layer from numerical
integrations of the equilibrium equation (\ref{e:1DSb}) for a flux
corresponding to the reference saturation degree $\hat{S}=0.30$ with
different downhill boundary-zone saturation degrees (labeled at
right).}
 \label{f:num30}
\end{figure}

Figure \ref{f:num30} presents equilibrium solutions to (\ref{e:1DSb})
formed by fourth-order Runge-Kutta numerical integration in the $+z$
upslope direction from a given fixed downhill saturation degree $S$
at $z=$ 20 km, as labeled on the right of each example, for the
reference saturation degree $\hat{S}=0.30$ and materials defined by
the van Genuchten coefficient combinations $\eta$ and $\psi_0$ given
with Figure \ref{f:vG}. The integration step of 1 m was chosen as the
largest step size that always produces stable and consistent results. 
The models illustrate the tendency for moisture to backup upslope at
near the saturation degree of the fixed downhill zone through a
characteristic pooling height determined by the boundary value and
material type.

\begin{figure}
\centering
\includegraphics[width=8.5cm,trim=10 50 -10 -50]{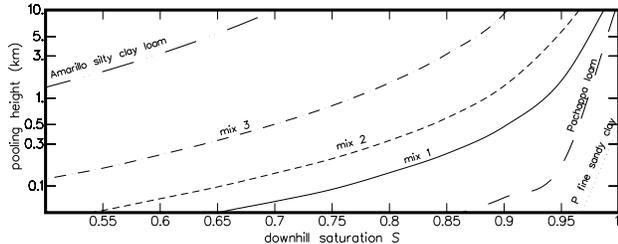}
 \caption{Pooling height (km) plotted on a log scale as a function of
the saturation degree $S$ in the downhill zone, synthesizing the
numerical experiments for the reference saturation degree
$\hat{S}=0.30$.}
 \label{f:poolht30}
\end{figure}

Figure \ref{f:poolht30} shows the pooling height as a function of the
boundary saturation degree, which is defined by locating the
intercept of the saturation curve above the lower boundary with the
upper-hillslope reference saturation line in numerical solutions as
from Figure \ref{f:num30}. The pooling height shows about a factor of
10 drop with every 10\% decrease in the downhill saturation degree
for any soil type near high saturation, though clays and silts
require less downhill saturation for a given pooling height.  The
pooling height is more than 10 km behind a downhill saturated zone in
all but the sandiest soil type, the Pachappa fine sandy clay.  For
average mix 1 or sandier, it decreases to less than 2 km with a
downhill saturation degree of $S=0.95$, and less than a km with
$S\leq0.9$. Though not shown, pooling heights decrease a little with
smaller reference saturation degree $\hat{S}$ and are not very
sensitive to the retained moisture saturation degree $S_{\rm r}$.

\section{Upper-Hillslope Flow Conditions}
\label{s:upper}

The Richards' Equation solution properties, the variations of
saturation degree and flow speed, are defined by the downslope flux.
The flux $q_\ell$ is defined by the product of the saturation degree
and flow speed in (\ref{e:Vdef}), or for the 1D solutions
 \begin{equation}
q_\ell=n S(\ell)V_\ell(\ell)=n\hat{S}\hat{V}_\ell.
\label{e:1DVdef}
\end{equation}
 The 1D equilibrium solutions from (\ref{e:1DSb}) determine the
downslope saturation function $S(\ell)$ for a given flux $q_\ell$ or
reference saturation degree $\hat{S}$.  Given a saturation function
$S(\ell)$, we obtain immediately the general product containing the
downslope flow speed $nV_\ell(\ell)$ using (\ref{e:1DVdef}), for what
may be possibly a downslope varying porosity $n(\ell)$. Since
$K(\hat{S})$ is a monotonically increasing function of $\hat{S}$ for
all material types from Figure \ref{f:vG}a, a constant reference
saturation degree $\hat{S}$ is uniquely defined by the constant flux
$q_\ell$ in (\ref{e:hatS}), so both the constant product with the
flow speed $n\hat{V}_\ell$ and constant saturation degree $\hat{S}$
for the upper hillslope are determined by the given flux $q_\ell$
using (\ref{e:hatS}) and (\ref{e:1DVdef}).

\begin{figure}
\centering
\includegraphics[width=8.5cm,trim=10 50 -10 -50]{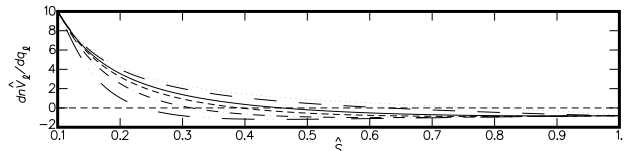}
 \caption{Sensitivity of upper-hillslope flow speed $\hat{V}_\ell$ to
changing flux $q_\ell$, $dn\hat{V}_\ell /dq_\ell$ as a function of
the reference saturation $\hat{S}$ for the van Genuchten materials
defined in Figure \ref{f:vG}.  To the right of the intercept with the
zero derivative line ({\it horizontal dashed line}), the flow speed
$\hat{V}_\ell$ decreases with increasing flux $q_\ell$.}
 \label{f:dVdq}
\end{figure}

The reference saturation degree $\hat{S}$ must always increase with
increasing flux $q_\ell$ from (\ref{e:hatS}).  However the
upper-hillslope flow speed in $n\hat{V}_\ell$ may either increase or
decrease with increasing $q_\ell$ from (\ref{e:1DVdef}).  The rate of
change $dn\hat{V}_\ell/d q_\ell$ can be determined by taking the
derivative $d/dq_\ell$ in (\ref{e:1DVdef}), which gives
 \begin{equation}
\hat{S}{dn\hat{V}_\ell\over dq_\ell}=1-n\hat{V}_\ell{d\hat{S}\over dq_\ell},
\end{equation}
 then substituting the derivative $d\hat{S}/dq_\ell$ from
(\ref{e:hatS}) and re-using (\ref{e:1DVdef}) and (\ref{e:hatS}) gives
 \begin{equation}
{dn\hat{V}_\ell\over dq_\ell}={1\over\hat{S}}
\left(1-{K(\hat{S})\over\hat{S}}\left({dK(\hat{S})\over d\hat{S}}\right)^{-1}\right).
\label{e:dVdq}
\end{equation}
 The discriminator $dn\hat{V}_\ell/dq_\ell$ is determined by the
hydraulic conductivity function $K(S)$ from (\ref{e:K}) alone and is
plotted in Figure \ref{f:dVdq} for all of the van Genuchten material
types defined in Figure \ref{f:vG}.  Where $dn\hat{V}_\ell /dq_\ell$
is negative, the flow speed in the upper hillslope decreases with
increasing flux.

\section{Flux Estimate}
\label{s:flux}

Our choice of reference saturation degree $\hat{S}=0.30$ for the La
Luz / Fresnal Watershed is reasonable considering the drop in outflow
seen at the upstream spring boxes, but seems justified too given
estimates of the precipitation inflow.  The constant flux down the
layer is defined by (\ref{e:hatS}).  For $\hat{S}=0.30$, $\iota=
7^\circ$, and the choice $K_{\rm sat}=7$ m/d, with Pachappa loam as
defined in Figure \ref{f:vG}, we obtain the flux $q_\ell=0.249$; with
mix 1, $q_\ell=0.326$ m/d; and with mix 2, $q_\ell=0.354$ m/d. 

The specific flux $q_\ell$ through the system is determined by the
total precipitation inflow normalized by the dimensions of the flow
layer
 \begin{equation}
q_\ell=\mu {Q_{\rm prec}\over D_y d}
=\mu \bar{q}_{\rm prec} \left({L_{\rm prec}\over d}\right),
\label{e:prec}
\end{equation}
 where $Q_{\rm prec}$ is the total precipitation inflow,
$\bar{q}_{\rm prec}= Q_{\rm prec}/(L_{\rm prec}D_y)$ is the specific
precipitation flux incident through the horizontal plane, $\mu$ the
relative precipitation fraction that goes into the downslope moisture
flow, $L_{\rm prec}$ the spatial extent of the high-precipitation
source region in $x$ above the flow layer, $D_y$ the spatial extent
of the source region into the plane of the drawing Figure
\ref{f:drawing} in $y$, and $d$ the thickness of the flow layer (see
Section 5.1, \citealt{Bear1988}). The results are independent of the
scale $D_y$, which divides out of the total precipitation inflow
$Q_{\rm prec}$.

For purposes of estimating the downslope flux, we suppose that the
total precipitation inflow $Q_{\rm prec}$ can be divided
systematically between components: ground moisture flow $Q_{\rm m}$,
streamflow $Q_{\rm sf}$, evapotranspiration loss $Q_{\rm et}$, and
immediate run off $Q_{\rm ro}$.  For the dry 2000s
 \begin{equation}
Q_{\rm prec}=Q_{\rm m}+Q_{\rm sf}+Q_{\rm et}+Q_{\rm ro},
\label{e:preca}
\end{equation}
 and for the wet 1980s
 \begin{equation}
\grave{Q}_{\rm prec}=\grave{Q}_{\rm m}+\grave{Q}_{\rm sf}+\grave{Q}_{\rm et}+\grave{Q}_{\rm ro}.
\label{e:precb}
\end{equation}
 The Tularosa Creek streamflow overresponse to the precipitation
change suggests that there was less relative evapotranspiration
during the wet period. Total evaporation is determined by the exposed
surface water, which should not change much with a small change in
precipitation. Although the transpiration physics is complicated,
natural mechanisms tend to compensate for decreased available
moisture in the root zone so that plant transpiration may remain
nearly constant with changing precipitation (\citealt{Salvucci2001};
\citealt{Teuling+++2006}), thus we take $\grave{Q}_{\rm et}= Q_{\rm
et}$. About 5\% leaves as immediate flood runoff in 3-7 large storm
events per year so $Q_{\rm ro}=0.05 Q_{\rm prec}$ and $\grave{Q}_{\rm
ro}=0.05 \grave{Q}_{\rm prec}$ (estimated using New Mexico Water
Resources Data, USGS yearly reports).  The highest precipitation from
1987 was 33\% higher than for the dry 2000s, $\grave{Q}_{\rm
prec}=1.33 Q_{\rm prec}$ from the average line in Figure
\ref{f:laluz}c, but the maximum relative Tularosa Creek streamflow
was actually larger $\grave{Q}_{\rm sf}=1.434 Q_{\rm sf}$ from Figure
\ref{f:laluz}b.

Though streamflow responds to the underground saturation in nonlinear
ways as we describe in Section \ref{s:conc}, supposing for overall
estimation purposes only that the streamflow stays approximately
proportional to the downslope moisture flow for the relatively small
changes as may be caused by precipitation variations in similar
downhill saturation conditions, we have $\grave{Q}_{\rm
m}+\grave{Q}_{\rm sf}=1.434(Q_{\rm m}+Q_{\rm sf})$. Using
$\grave{Q}_{\rm ro}=0.05 \grave{Q}_{\rm prec}= 0.05\cdot1.33 Q_{\rm
prec}=0.0665 Q_{\rm prec}$, we can eliminate all the quantities for
the wetter period in (\ref{e:precb}), giving
 \begin{equation}
1.2635 Q_{\rm prec}=1.434(Q_{\rm m}+Q_{\rm sf})+Q_{\rm et}.
\label{e:precc}
\end{equation}
 Then substituting $Q_{\rm ro}=0.05 Q_{\rm prec}$ in (\ref{e:preca})
and eliminating $Q_{\rm m}+Q_{\rm sf}$ with (\ref{e:precc}) gives the
relative evapotranspiration
 \begin{equation}
Q_{\rm et}=0.228 Q_{\rm prec},\ {\rm leaving}\ 
Q_{\rm m}+Q_{\rm sf}=0.722 Q_{\rm prec}.
\label{e:precd}
\end{equation}

Under normal downhill saturated conditions in the area system, about
15\% of the precipitation appears as streamflow so $Q_{\rm sf}=0.15
Q_{\rm prec}$, giving the ground moisture flow $Q_{\rm m}=0.572
Q_{\rm prec}$ for the moisture fraction $\mu= Q_{\rm m} / Q_{\rm
prec}=0.572$.  Under the reduced downhill saturation conditions in
the La Luz / Fresnal Watershed after 2002, the relative streamflow
was $Q_{\rm sf}=0.047 Q_{\rm prec}$, which gives an increased
relative moisture flow of $Q_{\rm m}=0.675 Q_{\rm prec}$ for
$\mu=0.675$.

The precipitation influx for the La Luz / Fresnal Watershed is
$\bar{q}_{\rm prec}=$ 25 in/yr = 1.74 mm/d asymptotically as a base
level for 2006 from Figure \ref{f:laluz}c.  We estimate the spatial
extent for the high-precipitation source region as $L_{\rm
prec}\simeq$ 12 km as the flattening top of the watershed above 7000
ft (2134 m) elevation partially shown in Figure \ref{f:lltopo}.  A
representative flow-layer thickness seems most uncertain, but we
assume a typical spacing of the impervious limestone strata of
$d\approx 40$ m characteristic of parts of the mountain system.  With
our guess and using the intermediate moisture fraction $\mu=0.624$,
we obtain $q_\ell=$ 0.326 m/d from (\ref{e:prec}) reasonably
consistent with the fluxes we obtained for the reference saturation
degree $\hat{S}=0.30$.  We must point out however that by adjusting
the parameters within acceptable ranges, we obtain a range of fluxes
and corresponding reference saturations.

The downslope moisture flow speed is defined $V_\ell= q_\ell/(nS)$
from (\ref{e:1DVdef}).  Taking a porosity $n=$ 0.17 for poorly sorted
soils, with the flux $q_\ell=0.326$ m/d and intermediate saturation
degree $S=0.6$, we obtain the flow speed $V_\ell=3.2$ m/d.

\section{Physical Interpretation}
\label{s:phys}

Steady equilibria like those derived in Section \ref{s:1Dsteady}
represent the long-term downslope profiles of saturation degree. Time
series as represented by (\ref{e:flux}) and (\ref{e:cont}), or by the
general Richards' Equation (\ref{e:flow}), or the 1D equations
(\ref{e:1Dflux}) and (\ref{e:1Dcont}) must tend to the nearest
equilibrium with a characteristic timescale determined by the flow
speeds for the problem.

The 1D steady lateral downslope equilibria are determined by two free
parameters: the flux $q_\ell$ and a boundary condition.  Streamflows
stay approximately uniform below the springs, but are absorbed into
the ground within a few km beyond the foot of the mountains shown on
the topo map in Figure \ref{f:lltopo}, where the geology changes from
sandy-clay limestone stratigraphy to alluvial fill.  There the
hydrology should change correspondingly from closed to open.  In the
basin, moisture pools with a variable deep water table and thus can
exert no pressure back upslope.  With an open boundary, no change in
the downhill saturation is imposed and the downslope flow must follow
the upper-hillslope form seen in all of the solution curves, a
constant flow speed $V_\ell= \hat{V}_\ell$ and saturation degree
$S=\hat{S}$ for the given flux $q_\ell$ from (\ref{e:1DVdef}), like
the solution reported in one study (\citealt{Philip1991a}).  With a
change of precipitation inflow for a new flux $q_\ell$, the flow
speed $\hat{V}_\ell$ and saturation degree $\hat{S}$ must adjust to
the new steady equilibrium defined in Section \ref{s:upper} with
Figure \ref{f:dVdq}.

At the onset of a new saturated zone, as may be produced with a
renewed sufficient continuous streamflow, the flux immediately
upslope $q_\ell$ defined in (\ref{e:1Dflux}) must be greatly
decreased and may even become negative by the imposed $\partial
S/\partial\ell$ positive singularity due to the large factor
$d\psi/dS$ near $S=1$. The downslope flux is slowed or may even
reverse its direction representing a net flux back upslope. The
reduced downslope flux produces an increased upslope saturation
according to the 1D continuity condition (\ref{e:1Dcont}).  In effect
the downslope moisture flow is blocked, and the blockage must
propagate back upslope as moisture piles up behind it leading to
equilibrium solutions (\ref{e:1DSb}) with profiles like those shown
in Figure \ref{f:num30}, which represent a return to the equilibrium
constant downslope-flux condition.  We may say that moisture backs up
behind the capillary head jump introduced in a zone of fixed
saturation degree through a pooling height determined by the balance
between the downslope gravity-driven flux out the bottom of the
blockage and the influx into the top.  If the saturated zone is not
sufficiently maintained, nonequilibrium or discontinuous behavior is
possible and the flow may again return to the reference saturation
below the saturated zone depending upon conditions further downslope.

During the transition from normal streamflow to no streamflow or from
a normally saturated downhill boundary condition to an open boundary
condition, a temporary increase in the total moisture outflow into
the basin below must occur.  The normal flux out into the basin will
return as the dryed system reaches a new equilibrium on the
adjustment timescale. The adjustment might be evidenced as a locally
increased water table in the basin, except that the Tularosa Basin is
so large that such changes may be difficult to detect and probably
could not be distinguished from other effects, like precipitation
change or distributional inhomogeneity.  The Tularosa Basin surface
area is about 13,500 km$^2$ compared to the La Luz / Fresnal
Watershed, which is about 200 km$^2$.  Also it seems that adequate
basin water-table measurements are not available.  

\section{Generalizations to 3D}
\label{s:3D}

In real 3D conditions, moisture can travel out of the mountain source
area along a number of independent paths, with different inflows,
downslope variations in conditions, and differing downhill saturation
zones on each path.  The 1D approximation should remain applicable in
the idealized flow layer of Figure \ref{f:drawing} even with changing
material properties downslope in $\ell$.  As the equilibrium
equations (\ref{e:1Dflux}), (\ref{e:K}), and (\ref{e:psi}), with
$q_\ell$ constant in $\ell$ are independent of the porosity $n$, a
changing porosity downslope does not affect the downslope saturation
solutions, but does affect the flow speed $V_\ell(\ell)$ through
(\ref{e:1DVdef}).  A changing hydraulic conductivity or capillary
pressure downslope in $\ell$ does not affect the 1D assumptions
either since it only changes the material functions that enter
(\ref{e:1DSb}), so solutions for $S(\ell)$ should be like what is
found in the uniform cases.

A slowly changing downslope flux $q_\ell$, as produced with an added
precipitation influx, or evapotranspiration or flow-layer losses, in
a thin flow layer of constant cross section must produce a slightly
changed saturation degree downslope $S(\ell)$ in the integrated
(\ref{e:1DSb}), but should not alter the essential solution
properties.  Such variations should produce systematic modifications
to the pooling heights above saturated zones.  

With variations in conditions across the flow layer or a varying
cross section or inclination downslope, non-lateral components to the
flux vector may be introduced and the 1D approximation break down,
consistent with the belief that underground geological and
topographical structure gives rise to deeper saturated structure like
perched aquifers (\citealt{JacksonCR1992};
\citealt{Neuweiler+Cirpka2005}).  In principle, the four equilibrium
equations from (\ref{e:cont}) ${\vec\nabla}\cdot{\vec q}=0$, and the
three components of (\ref{e:flux}) define the three components of
${\vec q}$ and $S$ throughout the 3D volume of the flow layer.  The
$q_\ell$ flux equation (\ref{e:1Dflux}) still represents the
downslope component in the general 3D conditions and must be
preserved in integrated form in the equilibrium solutions.  Thus we
argue that in 3D conditions where a nonnegligible downslope flux
covering the cross section exists down the flow layer, the basic
physical conservation principles from the 1D solutions should still
apply, and flow backup similarly above downhill saturated zones that
cover the cross section of a closed flow layer. 
\cite{Hurley+Pantelis1985} define special forms of variations that
still preserve the 1D flow vector.  

The 1D approximation breaks down too in the boundary zone if the
saturating process by streamflow produces a non-uniform moisture
distribution underground. Variations in the upslope pooling
properties can occur if the underground is inhomogeneously partially
saturated rather than being uniformly filled with a single saturation
degree, as idealized in our 1D solutions.  While the two cases of a
fully saturated boundary zone covering the cross section of the flow
layer with maximum pooling above and an open boundary with no pooling
are still correctly represented, partial coverage may change the
properties of the downslope flow blockage and modify the pooling
heights between the two extremes.

\section{Conclusions}
\label{s:conc}

The hypothesis that water leaves the mountains mainly in a
large-scale near-surface moisture flow is idealized by our 1D
Richards' Equation model for lateral downslope steady flow in a
uniformly filled thin closed layer of constant cross section and
inclination. The Richards' Equation represents general
saturated/unsaturated underground moisture dynamics.  The 1D steady
flow equation is directly integratable and exhibits a constant
downslope flow speed and moisture content determined by the constant
downslope flux.  At a boundary zone of increased saturation degree,
the flow speed is slowed and moisture content increased back upslope
through a pooling height, characteristized by the equilibrium
condition that the flux, the product of the flow speed and moisture
content, is constant. Surface streamflow must produce zones of
increased saturation degree that affect the underground saturated
boundary condition.  Such a strong solution dependence on the form of
the lower boundary condition suggests that mountain moisture content
in hillslopes above deep alluvial basins may be strongly affected by
lower streamflow.

In our numerical integrations, the pooling height is more than 10 km
vertically projected behind a fully saturated zone in all but the
sandiest soil types, but decreases rapidly with a decreasing boundary
saturation degree.  The other cited studies of downslope moisture
flow in a thin layer also show a constant flow speed and saturation
degree or pooling at the bottom asymptotically in their time series,
but our pooling heights appear to be much larger than what is found
in the other studies.  The large scale arises mainly because material
capillary head functions exhibit singular behavior near full
saturation, and all of the cited studies that show downstream pooling
use different approximations for the material hydraulic conductivity
or capillary-head functions to obtain a more tractable problem.

Spring outflow must be a direct result of the coalescence of
underground moisture in irregular topography near or upslope from the
spring (\citealt{Freeze1972b}; \citealt{Fipps+Skaggs1989}).  With a
reduced downhill saturation, the saturation degree outside the
streambed and back up to the spring may be greatly decreased from a
near 100\% pooling-determined saturation degree to a 30\% reference
saturation degree for representative conditions in the La Luz /
Fresnal Watershed.  Thus the drop of $~69\%$ in spring outflow, as
the unaccounted loss seen at Alamogordo's upstream spring-box
diversions, seems a reasonable possibility for the downhill drying,
taking the spring outflow to be simply proportional to the saturation
degree in the spring locale.  Though individual spring response to
changing conditions may be nonlinear, without knowing the specifics,
the general suggestion from studies is that the central hillslope
outflow increases with the local water-table height (cited references
and \citealt{Weyman1973}).  The mountain saturation above may be
significantly, but probably less compromised by downhill drying, as
presumably not all of the moisture paths out of the mountain source
area are affected.  Also the adjustment time for the mountains above
is longer and effects there should become evident only more slowly.

The rate at which moisture moves downslope from a disappearing
saturated zone or backs up behind a newly established one might be
estimated supposing that an unmaintained saturated zone will
dissipate downslope at the average flow speed, or flow backup behind
a newly formed saturated zone at the same rate.  The actual timescale
for depletion of the La Luz / Fresnal Watershed seen in Figure
\ref{f:laluz}a and known from the history of the Alamogordo project
was about $3-10$ years, which for the flow speed $V_\ell=3.2$ m/d is
the time required to travel $3.5-11.7$ km, not out of range for the
spring distances from the mountain-basin outlet shown in Figure
\ref{f:lltopo}.  The estimate is consistent with the saturated-flow
estimate described at the end of Section \ref{s:1Dsteady}.  We note
however the large uncertainty in the flow speed due especially to the
uncertainties in the saturated hydraulic conductivity in real field
conditions.


\end{article}
\end{document}